\documentclass[12pt]{article}
\usepackage{amsmath,amssymb,amsfonts,epsfig,graphicx,euscript}
\newcommand{\be}{\begin{equation}}
\newcommand{\ee}{\end{equation}}

\newcommand{\bse}{\begin{subequations}}
\newcommand{\ese}{\end{subequations}}
\newcommand{\bea}{\begin{eqnarray}}
\newcommand{\eea}{\end{eqnarray}}
\newcommand{\ba}{\begin{array}}
\newcommand{\ea}{\end{array}}

\makeatletter \@addtoreset{equation}{section}

\makeatother

\begin{document}
\baselineskip 18pt%


\begin{center}
{ \Large{\textbf{On meson melting in the quark medium }
\\}}
\vspace*{2cm}
\begin{center}
{\bf K. Bitaghsir Fadafan$^1$ and E. Azimfard$^2$}\\%
\vspace*{0.4cm}
{\it {$^{1,2}$Physics Department, Shahrood University of Technology,\\
P.O.Box 3619995161, Shahrood, Iran }}\\
{E-mails: {\tt bitaghsir@shahroodut.ac.ir, ehsan.azimfard@gmail.com}}%
\vspace*{1.5cm}
\end{center}
\end{center}

\vspace{.5cm}
\bigskip
\begin{abstract}
We consider a heavy quark-antiquark $(q\bar{q})$ pair as a heavy
meson in the medium composed of light quarks and gluons. By using
the AdS/CFT correspondence, the properties of this system are
investigated. In particular, we study the inter-quark distance and
it is shown that the mechanism of melting in the quark-gluon plasma
and in the hadronic phase are the same. It is found that by
considering finite coupling corrections, the inter-quark distance of
a heavy meson decreases. As a result a heavy meson like $J/ \psi$
will melt at higher temperatures. By considering rotating heavy
mesons, we discuss melting of exited states like $\chi_c$ and
$\psi'$.

\end{abstract}

\newpage
\tableofcontents

\section{Introduction}
The experiments of Relativistic Heavy Ion Collisions (RHIC) have
produced a strongly-coupled quark$-$gluon plasma (QGP)(see review
\cite{CasalderreySolana:2011us}). At a qualitative level, the data
indicate that the QGP produced at the LHC is comparably
strongly-coupled \cite{ALICE}. The QGP that is created at the LHC is
expected to be better approximated as conformal than is the case at
RHIC \cite{CasalderreySolana:2011us}. There are no known
quantitative methods to study strong coupling phenomena in QCD which
are not visible in perturbation theory (except by lattice
simulations). A new method for studying different aspects of QGP is
the $AdS/CFT$ correspondence
\cite{CasalderreySolana:2011us,Maldacena:1997re,Gubser:1998bc,Witten:1998qj,
Witten:1998zw}. This method has yielded many important insights into
the dynamics of strongly-coupled gauge theories. It has been used to
investigate hydrodynamical transport quantities in various
interesting strongly-coupled gauge theories where perturbation
theory is not applicable. Methods based on $AdS/CFT$ relate gravity
in $AdS_5$ space to the conformal field theory on the
four-dimensional boundary \cite{Witten:1998qj}. It was shown that an
$AdS$ space with a black brane is dual to a conformal field theory
at finite temperature \cite{Witten:1998zw}.

In heavy ion collisions at the LHC, heavy-quark related observables
are becoming increasingly important \cite{Laine:2011xr}. In these
collisions, one of the main experimental signatures of QGP formation
is melting of heavy mesons in the medium \cite{Matsui:1986dk}. In
this paper we study melting of a heavy meson from AdS/CFT. We
consider a heavy quark-antiquark $(q\bar{q})$ pair in the medium
composed of light quarks and gluons as a heavy meson
\cite{Erdmenger:2007cm}. By using the AdS/CFT correspondence, the
properties of this system are investigated.

We will consider the holographic QCD in the quark medium which was
studied in \cite{Lee:2009bya,Kim:2007em}. It was shown that at the
high temperature, the gravity dual to the quark-gluon plasma is the
Reissner-Nordstrom AdS $(RNAdS)$ black hole and at the low
temperature, the dual geometry corresponding to the hadronic phase
is the thermal charged AdS $(tcAdS)$ space. It was also found that
$(tcAdS)$ space can be obtained by taking zero mass limit of the
$RNAdS$ black hole \cite{Lee:2009bya}. The confinement/deconfinement
phase transition in the quark medium was discussed in
\cite{Kim:2007em} and an influence of matters on the deconfinement
temperature, $T_c$ was investigated. Using a different normalization
for the bulk gauge field, it was shown that the critical baryonic
chemical potential becomes $1100 MeV$ which is comparable to the QCD
result \cite{Park:2009nb}.

Melting of a heavy meson is investigated in \cite{Park:2009nb} and
it is found that in the quark-gluon plasma the dissociation length
\footnote{One may call it screening length.} becomes shorter as the
chemical potential increases. On the contrary, in the hadronic phase
the dissociation length becomes larger as the chemical potential
increases. In the next section, we argue that this conclusion is not
valid and explicitly show that in the hadronic phase (low
temperature), the dissociation length decreases by increasing the
chemical potential. We conclude that the melting mechanism in the
quark-gluon plasma and in the hadronic phase are the same
{\em{i.e}}. the interaction between heavy quarks is screened by the
light quarks.

In third section, we consider higher derivative corrections
{\em{i.e}}. ${\cal{R}}^2$ which corresponds to the finite coupling
corrections on the \emph{static} quark-antiquark system in the hot
plasma. To study ${\cal{R}}^2$ corrections, static heavy meson in
Gauss-Bonnet background has been considered. It is shown that at a
given energy, the increase in the $\lambda_{GB}$ leads to a decrease
in the dissociation length. This confirms the results of
\cite{Fadafan:2011gm} {\em{i.e}}. considering the higher derivative
corrections in the gravity background leads to a decrease in the
dissociation length of heavy meson in the boundary gauge theory. One
may conclude that the heavy meson melting at finite coupling occurs
at higher temperature which confirms that heavy mesons like
quarkonia need higher temperature to melt. It was explored in
\cite{Matsui:1986dk} that $J/ \psi$'s can be used as a
confinement/deconfinement probe. As a result, $J/\psi$ will melt at
higher temperature than the excited states like $\chi_c$ and $\psi'$
\cite{Satz:2009hr}.

To study melting of excited states in the quark medium, we consider
a rotating heavy meson. It is shown that the dissociation length
increases as the spin increases which confirms that the excited
states will melt at lower temperature than the ground states. We
also investigate the effect of increasing flavor quarks on the
dissociation length. It is shown that as the flavor quarks increase,
the dissociation length decreases. We summarize the results in the
conclusion section.

\section{Holographic melting in the quark medium}
In this section we will give a brief review of \cite{Park:2009nb}.
The Euclidean action describing the five-dimensional asymptotic $
AdS$ space with the gauge field is given by%

\be S = \int d^5 x \sqrt{G} \left( \frac{1}{2 \kappa^2} \left( -
{\cal R} + 2 \Lambda\right)  + \frac{1}{4g^2} F_{MN} F^{MN} \right)
, \label{S1}\ee%
where $\kappa^2$ is proportional to the five-dimensional Newton
constant and $g^2$ is a five-dimensional gauge coupling constant.
The cosmological constant is given by $\Lambda = \frac{-6}{R^2}$,
where $R$ is the radius of the $AdS$ space. It was pointed out that
the quark-gluon plasma and hadronic phase could be described by the
$tcAdS$ and the $RNAdS$ black hole, respectively \cite{Park:2009nb}. These solutions can be considered as follows%

\be ds^2= \frac{R^2}{z^2} \left(  f(z) dt^2 + d \vec{x}^{ 2} +
\frac{1}{f(z)} dz^2 \right)  ,
\label{ds2}\ee%
where the function $f(z)$ for $RNAdS$ black hole is given by%

\be f(z)=1-mz^4+q^2z^6,\label{fQGP}\ee%
and in the case of $tcAdS$ space%
\be f(z)=1+q^2z^6.\label{fhadronic}\ee%

In these coordinates, $z$ denotes the radial coordinate of the black
hole geometry and $t, \vec{x}$ label the directions along the
boundary at the spatial infinity. The event horizon is located at
$f(z_h)=0$ where $z_h=z_+$ is the largest root and it can be found
by solving this equation. The boundary is located at $z=0$ and the
geometry is asymptotically $AdS$ with radius $R$.

The parameters $ m$ and $q$ are the black hole mass and charge,
respectively. The time-component of the bulk gauge field is
$A_t(z)=i(2\pi^2\mu-Q\,z^2)$ where $\mu$ and $Q$ are related to the
chemical potential and quark number density in the dual gauge
theory. Regarding the Drichlet boundary condition at the horizon,
$A_t(z_+)=0$, one finds $Q=\frac{2 \pi^2 \mu}{z_+}$. The black hole
charge $q$ and the quark number density $Q$ also
are related to each other by this equation%
\be Q=\sqrt{\frac{3g^2R^2}{2\kappa^2}}\,q. \ee%

From now on, we investigate the holographic melting of heavy mesons
in two phase of the quark medium, the quark-gluon plasma and  the
hadronic phase, respectively. The holographic melting of heavy
mesons at finite temperature in a chiral and confining string dual
is discussed in \cite{Peeters}. It is argued in \cite{Rey} that the
heavy quark and anti-quark potential becomes zero at a separation
$r_d$ and for large separation, the dominant configuration
corresponds to two straight strings and the string melts. Then, one
should study the binding energy of heavy meson $V_{b}(r)$ when it
goes to zero\footnote{This is the large $N_c$ and large 't Hoof
coupling definition of heavy quark potential. A careful discussion
of the corrections to this definition was done in
\cite{Bak:2007fk}.}. This phenomena happens at special length $r_d$
which is obtained from $ V_{b}(r_d)=0 $. One may call $r_d$ as the
screening length of heavy meson \cite{Antipin}. The screening length
$L_s$ of a heavy quark-antiquark pair in strongly coupled gauge
theory plasmas flowing at velocity $v$ is studied in
\cite{Liu:2006nn}. There, screening length is defined so that for
$L>L_s$ no extremal world-sheet exists which binds the
quark-antiquark pair. Regarding the conventions of
\cite{Park:2009nb}, we call $r_d$ dissociation length.

One finds from the standard calculations
of \cite{Park:2009nb} that the inter-quark distance is given by %
\be r=2 \int_0^{z_0} dz\, z^2\,\,\frac{\sqrt{f(z_0)}}{\sqrt{f(z)}}\,
\frac{1}{\sqrt{f(z)z_0^4-f(z_0)z^4}}\,.\label{r}
\ee%
The inter-quark distance in AdS-Schwarzschild black hole is
investigated in \cite{Rey}. It is shown that for long strings
\footnote{We would like to thank S. Sheikh-Jabbari for discussion on this expansion.}%
\be \frac{r}{2}= A\, c\left(\frac{1}{a}-\frac{1}{5a^5}-\frac{1}{10a^9}-...\right)\label{expansion}\ee%
where $A$ is a constant and%
\be c=\sqrt{2}\,
\mathbf{E}(1/ \sqrt{2})-\frac{1}{\sqrt{2}}\,\mathbf{K}(1/\sqrt{2}).\label{c}\ee%

The inter-quark distance of a static meson in the quark-gluon phase
and in two special limits can studied as follows
\begin{itemize}
\item{$z_0\simeq z_+$ \\%
This limit correspond to large strings which means that the U-shaped
string touches the horizon. Then one should assume $ f(z_0)=\epsilon
\simeq 0$ and the equation for
$r$ becomes%
\be r=\frac{2\sqrt{\epsilon}}{z_0^2} \int_0^{z_0} dz\,
\,\frac{z^2}{f(z)}.\label{rlimitup}
\ee%
It is clearly seen that there is singularity which is reasonable. In
the case of extremal $RNAdS$ black hole, one finds that $f(z) \simeq
(z-z_0)^2$ .
}
\item{$z_0\simeq 0$\\%
Now we consider $z_0\rightarrow \epsilon \simeq 0$. It correspond to
the short strings. In this case
$f(\epsilon)\simeq 1$ and one finds that \eqref{r} becomes %
\be r/2=\int_0^{\epsilon}\frac{dz}{\sqrt{-q^2 z^6+m z^4-1}}\label{rz00}\ee%
where the solution is expressed in terms of Incomplete Elliptic
functions. See appendix \textbf{A}.%
}
\end{itemize}
\begin{figure}[ht]
\centerline{ \includegraphics[width=3in]{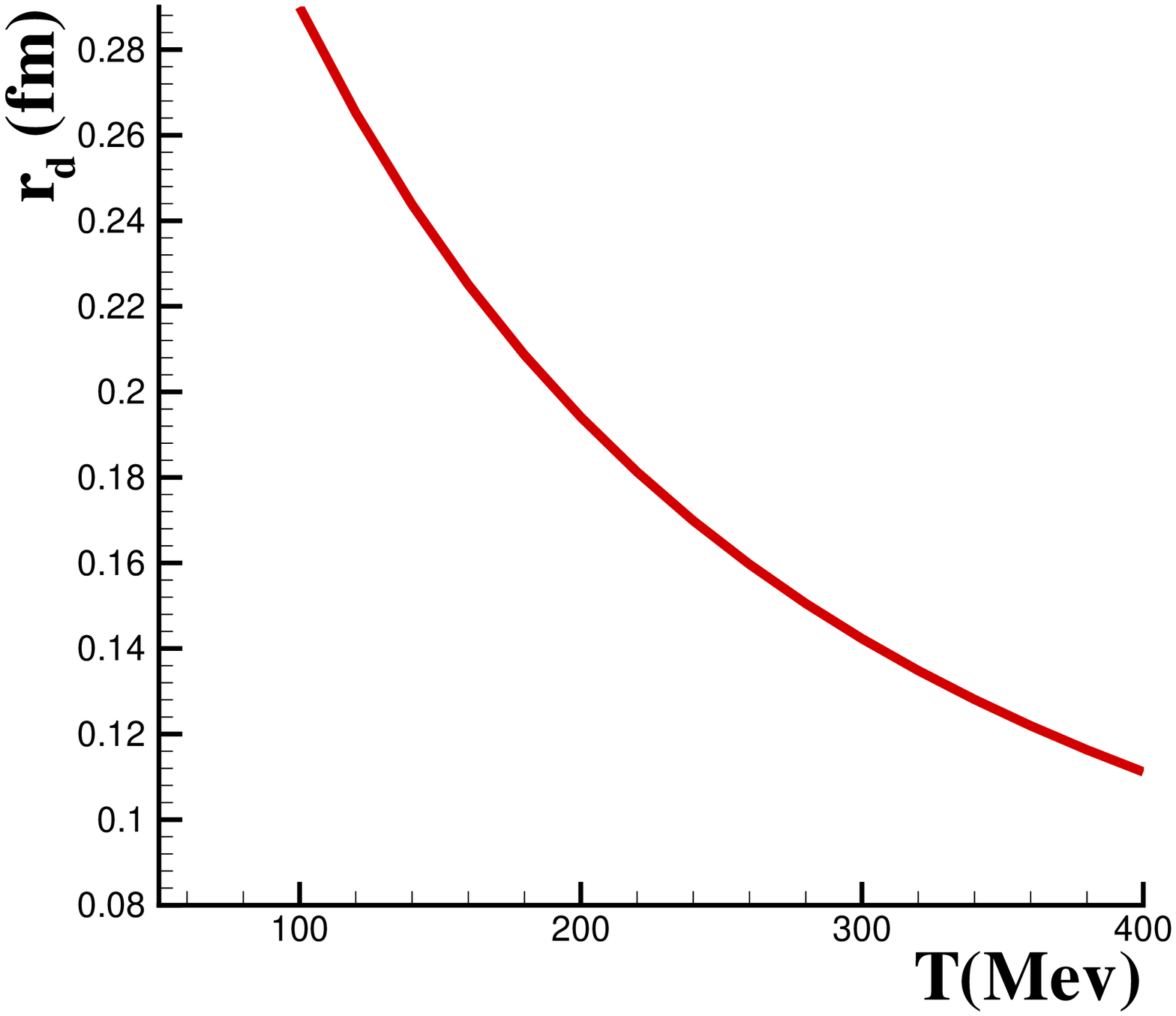},\includegraphics[width=3in]{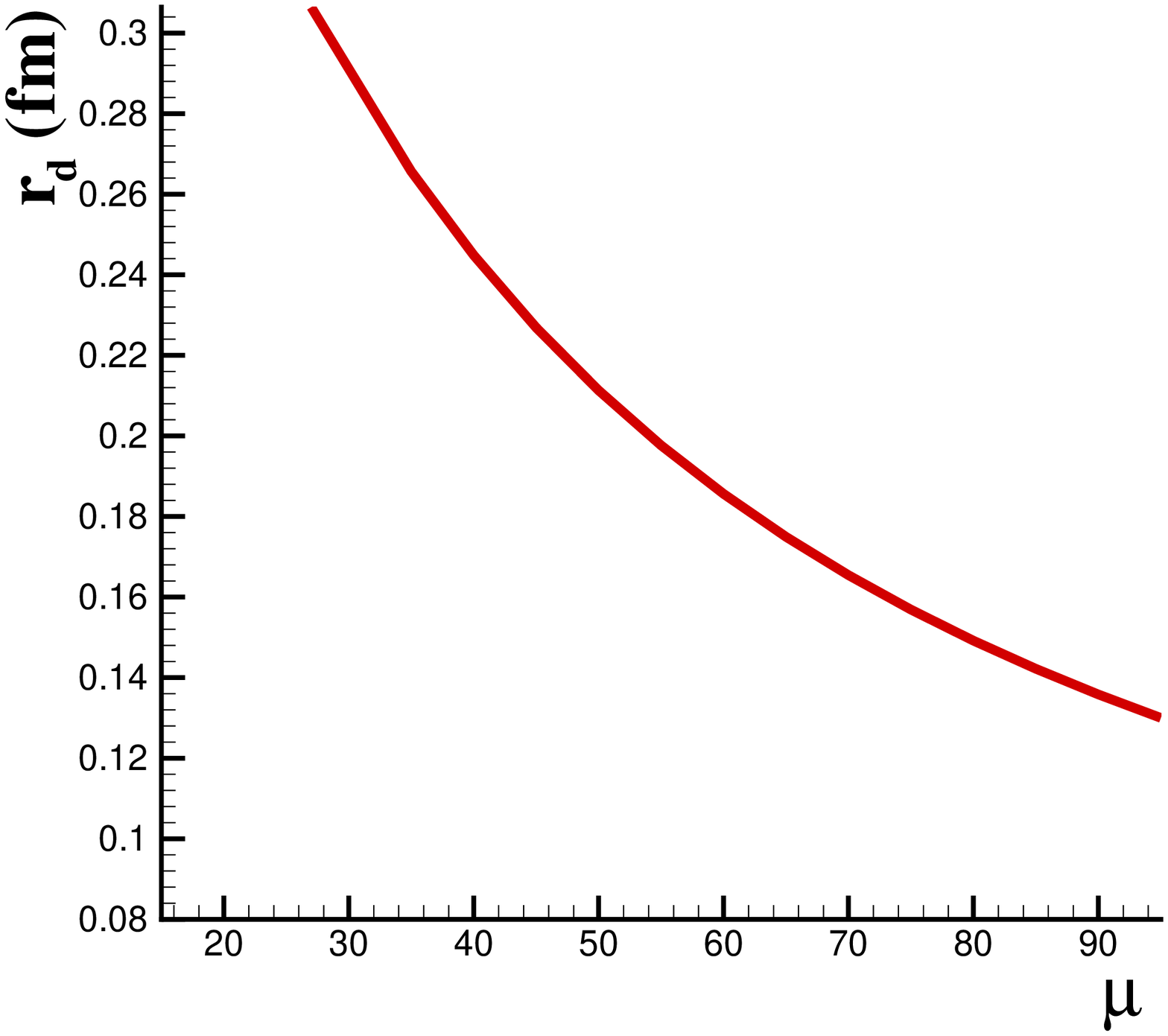}} %
\caption{Left: The inter-quark distance versus the temperature of
the quark-gluon plasma at fixed chemical potential $\mu=30 MeV$ and
$n=1$. Notice that the temperature must be larger than $91.2 MeV$.
Right: The inter-quark distance versus the chemical potential of the
quark-gluon plasma at fixed temperature $T=100 MeV$ and $n=1$.
Notice that $\mu>26.8 MeV$. }
\end{figure}%
One finds from the standard calculations
of \cite{Park:2009nb} that the binding energy is %
\be V_b=\frac{R^2}{\pi \alpha '} \left( \int_0^{z_0} dz\,
z^{-2}\,\,\,
\frac{\sqrt{f(z)}}{\sqrt{f(z)-\frac{f(z_0)z^4}{z_0^4}}}-\int_0^{z_1}
dz\, z^{-2}\right)\label{V}
\ee%

In the above equation, the second term is the energy for two free
heavy quarks.\footnote{The main mistake in the calculations of
\cite{Park:2009nb} comes from this term. This mitake has an
important effect in the hadronic phase when we study the
dissociation length versus the chemical potential in Fig. 2.} The
nearest point of U-shaped string to the horizon is shown by $z_0$.
Notice that $z_1>z_0$ and in the quark-gluon plasma $z_1$ is the
outer horizon of the $RNAdS$
black hole $z_+$ which is given by%
\be z_+=\frac{3g^2R^2}{8\pi^4\kappa^2\mu^2}\left(\sqrt{\pi^2T^2+
\frac{16 \pi^4\kappa^2\mu^2}{3g^2R^2}}-\pi T\right).\ee%
where $T$ is the temperature of the quark-gluon plasma. In the
hadronic phase, $z_1$ is the IR cut-off $z_{IR}$ \cite{Park:2009nb}.

We repeat the calculations of \cite{Park:2009nb} and plot the $r_d$
depending on the temperature and the chemical potential in Fig. 1.
This figure corresponds to Fig. 3. in \cite{Park:2009nb}. These
plots imply that the $r_d$ becomes short as the temperature and
chemical potential of quark-gluon plasma increases.
\begin{figure}[ht]
\centerline{\includegraphics[width=3in]{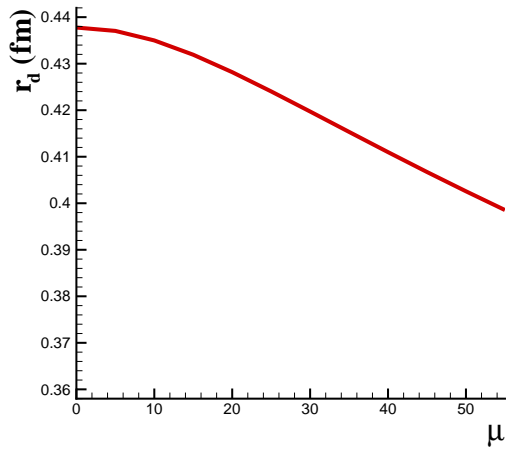}}%
\caption{The dissociation length versus the chemical potential in
the hadronic phase and $n=1$. This phase exists at the zero
temperature then we do not have the temperature dependency of the
 $r_d$. }
\end{figure}%
In the hadronic phase, one should notice
 that in \eqref{r} and \eqref{V}, $f(z)$ is given by \eqref{fhadronic}.
 It would be interesting to investigate the behavior of $r_d$ in terms of the chemical potential.
 We show the result in Fig.
 2. It is clearly seen that by increasing the chemical potential,
 the dissociation length becomes shorter. On the contrary, the Fig. 5 of \cite{Park:2009nb} shows that the dissociation length becomes larger.
 We found that the reason of this wrong result is the mistake in the free energy of two quarks in
 \eqref{V}.\footnote{We would like to thank Chanyong Park for discussion on this point.}
  As a result one concludes similar physics for melting of
 heavy mesons in the quark-gluon plasma and hadronic phase of QCD. In the QGP
 phase, the interaction between heavy quarks is screened
by the light quarks so that the dissociation length of the heavy
meson decreases as the temperature or the light quark chemical
potential becomes larger.

It is clearly seen that the shape of $r_d$ versus the chemical
potential in the quark-gluon plasma and in the hadronic phase are
not the same. Compared with the Fig. 2, the decrease in dissociation
length in Fig. 1 is more noticeable. One possible reason is that in
the hadronic phase, there is no temperature dependence in the
holographic QCD model and one considers the zero temperature case
only. As a result, the role of temperature in the melting of heavy
meson is absent in the hadronic phase.
\begin{figure}[ht]
\centerline{\includegraphics[width=3in]{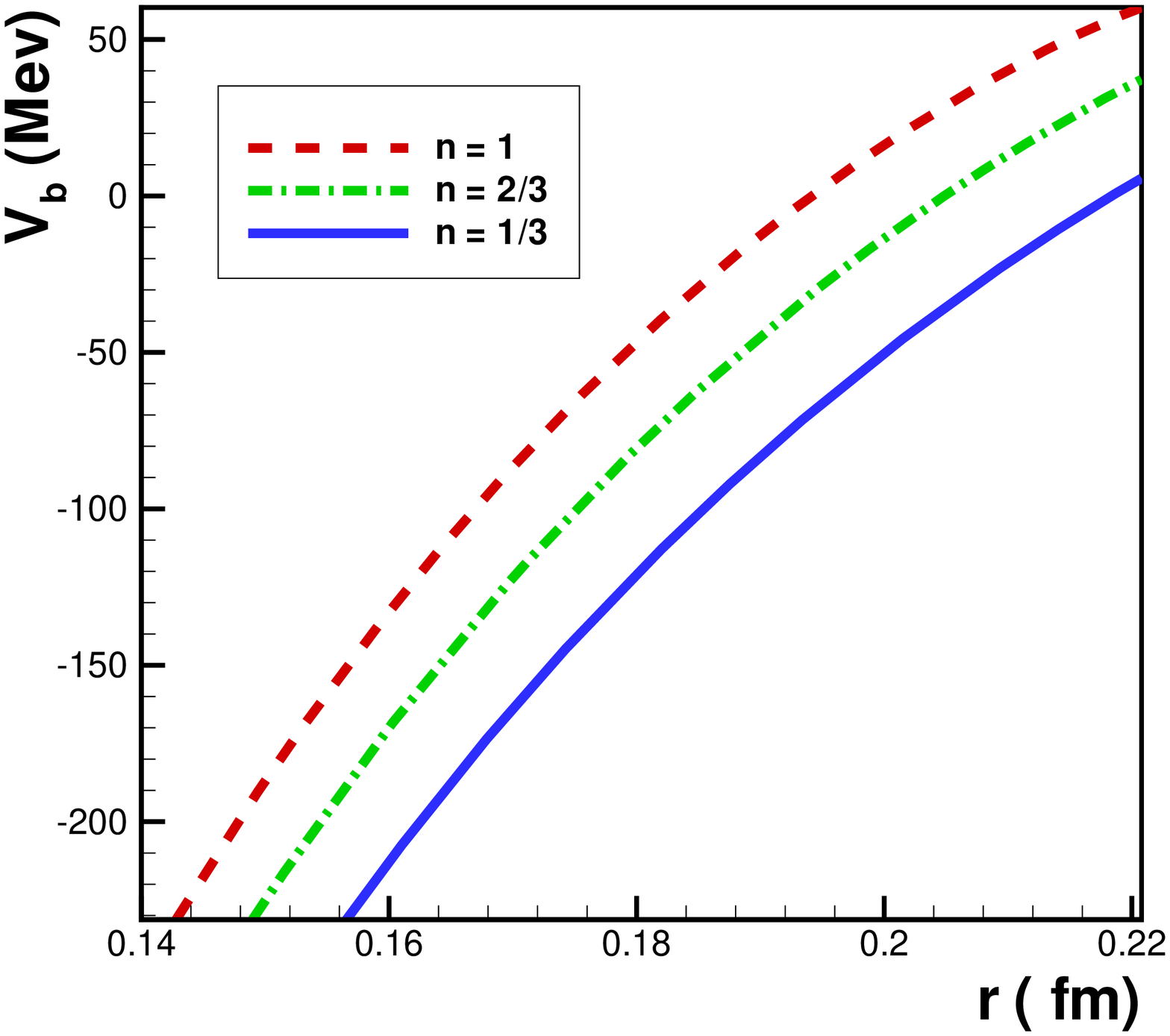}\includegraphics[width=3in]{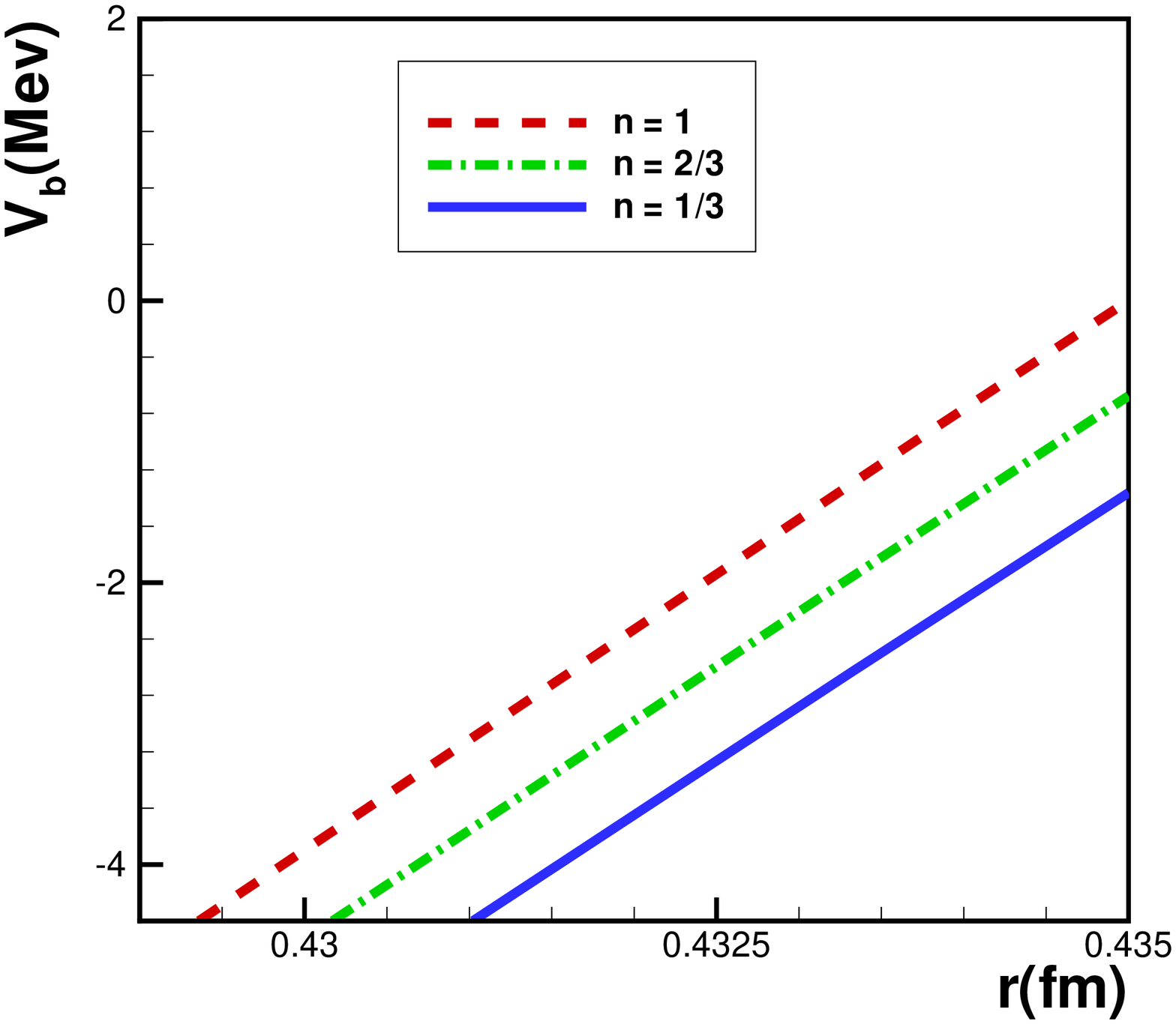}}%
\caption{Left: The binding energy versus the inter-quark distance in
the quark-gluon plasma at $(\mu,T)=(30\,MeV,  200\,MeV)$. Right: The
binding energy versus the inter-quark distance in the hadronic phase
and $(\mu,T)=(10\,MeV,  0\,MeV)$. }
\end{figure}%

We introduce $N_f=nN_c$ where  $N_f$ and $N_c$ are the number of
flavors and color fields, respectively. We are going to study how
$r_d$ changes when flavor number of quarks increases
\cite{Park:2009nb}. We plot the binding energy of heavy meson versus
the inter-quark distance in the Fig. 3. In the left and right plot
of this figure, we have considered the quark-gluon plasma and
hadronic phase of QCD. It is clearly seen that in each phase of QCD,
$r_d$ becomes shorter as the flavor number becomes larger. Compared
with the Fig. 4. of \cite{Park:2009nb}, one finds that in the
hadronic phase as the flavor number increases $r_d$ decreases. This
implies that as the flavor number increases, the heavy meson melts
easier. The meaning of the melting in the hadronic phase was
discussed in \cite{Park:2009nb}. It implies that the heavy meson is
broken into the light mesons which forms from two heavy-light quarks
bound states.

\section{Holographic melting at finite coupling}
In this section, we investigate the effect of the curvature-squared
corrections on the screening length. These corrections are related
to the finite-coupling correction in the corresponding gauge theory.
It was shown that these corrections affect the dissociation length
and it was argued that the increase of the coupling of the
corrections decreases the $r_d$ \cite{Fadafan:2011gm}. Now we
continue our study to the case of a heavy meson in the quark medium.

To study effects of the finite 't Hooft coupling, we consider the
Reissner-Nordstr\"{o}m-AdS black brane solution in Gauss-Bonnet
gravity \cite{Cvetic:2001bk}. The following action term in 5
dimensions describes the Gauss-Bonnet term which should be
considered in \eqref{S1} as
\begin{figure}[ht]
\centerline{\includegraphics[width=3in]{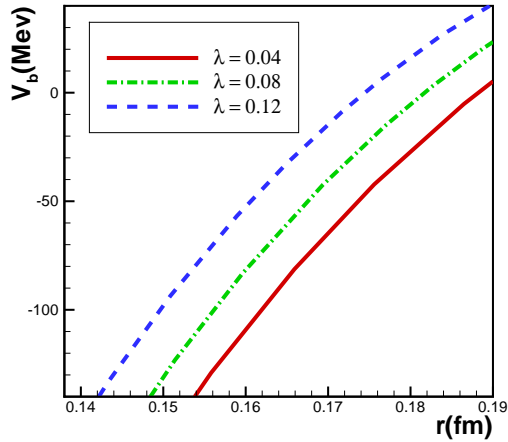}}%
\caption{The binding energy of the heavy meson versus the
inter-quark distance for different values of $\lambda= 0.04, 0.08$
and $0.12$. Graph is made for $n=1$ and $(\mu, T)=(30\,MeV ,
200\,MeV).$}
\end{figure}%
\begin{eqnarray}
S_{\mathcal{R}^2}=\frac{R^2\,\lambda_{}}{4\kappa^2}\left(
\mathcal{R}^2-4 \mathcal{R}_{\mu\nu}\mathcal{R}^{\mu\nu}+
\mathcal{R}_{\mu\nu\rho\sigma}\mathcal{R}^{\mu\nu\rho\sigma}\right)\label{GBFaction},
\end{eqnarray}
where $\mathcal{R}_{\mu\nu\rho\sigma}\,,\mathcal{R}_{\mu\nu}$ and
$\mathcal{R}$ are the Riemann curvature tensor, Ricci tensor, and
the Ricci scalar, respectively. The Gauss-Bonnet coupling constant
is $\lambda_{}$. The charged black brane solution in 5-dim is given
by%
\be ds^2= \frac{R^2}{z^2} \left(  N^2 f(z) dt^2 + d \vec{x}^{ 2} +
\frac{1}{f(z)} dz^2 \right)  ,
\label{ds2GB}\ee%
where
\begin{equation}
f(z)= \frac{1}{2\lambda_{}}\left( 1-\sqrt{1-4 \lambda_{}\left(
1-mz^4+q^2z^6 \right)}\right).
\end{equation}

The constant $N^2$ is arbitrary which specifies the speed of light
of the boundary gauge theory and we choose it to be unity. As a
result at the boundary, where $z\rightarrow 0$,
\begin{equation}
f(z)\rightarrow \frac{1}{N^2 }, \,\,\,\,\, N^2= \frac{1}{2}\left(
1+\sqrt{1-4 \lambda_{}} \right)\label{a}.
\end{equation}
Beyond  $\lambda_{}\leq\frac{1}{4}$ there is no vacuum AdS solution
and one cannot have a conformal field theory at the boundary.
Casuality leads to new bounds on $\lambda_{}$
\cite{Brigante008gz,Ge:2009eh}. The drag force on a moving heavy
quark and the jet quenching parameter in the background of $RNAdS$
black hole in Gauss-Bonnet gravity was studied in
\cite{Fadafan:2008uv}. By setting Gauss-Bonnet coupling to be zero,
the analytic solution for drag force in the case of
Reissner-Nordstr\"{o}m-AdS background was found.

\begin{figure}[ht]
\centerline{\includegraphics[width=3in]{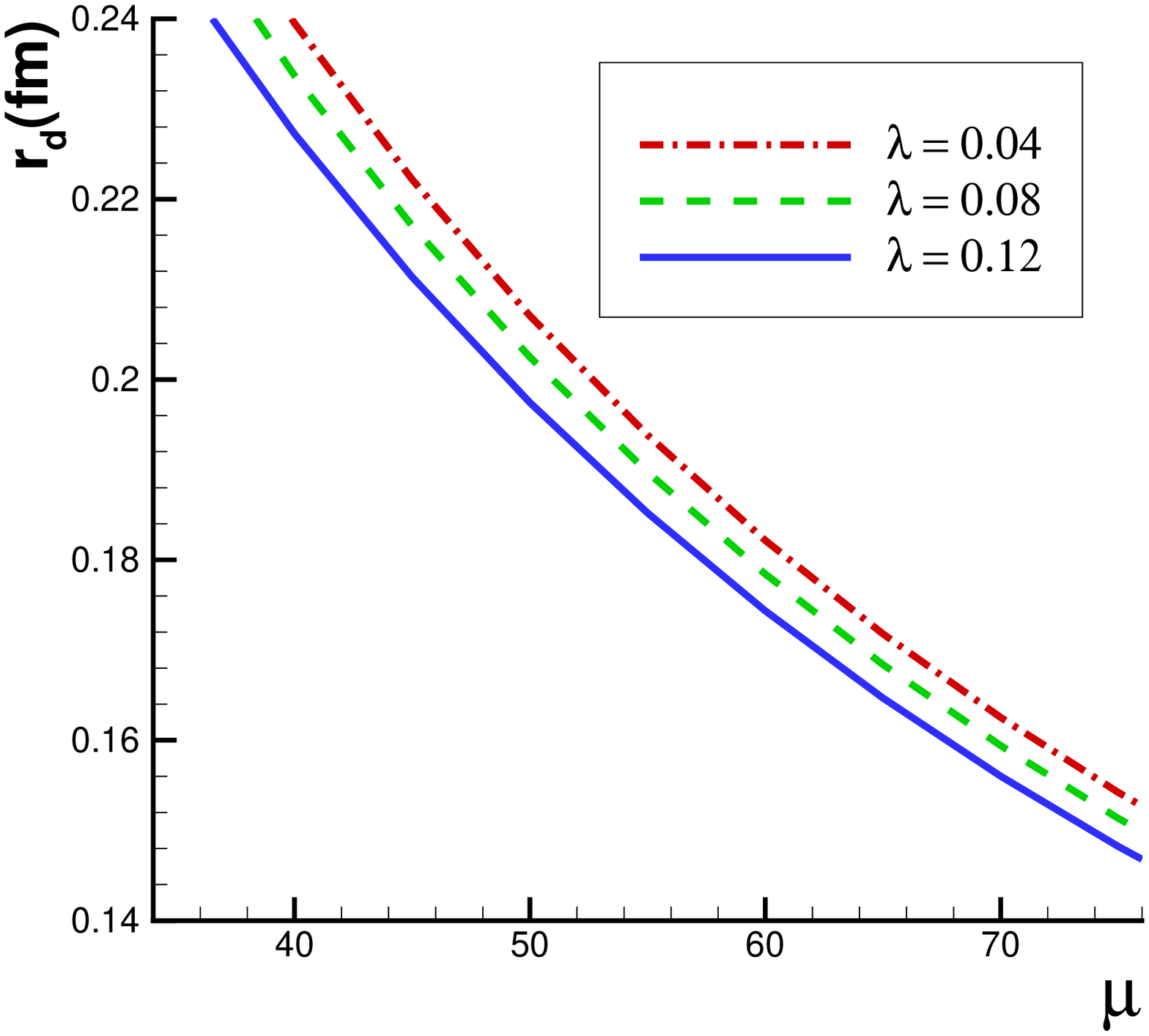}\includegraphics[width=3in]{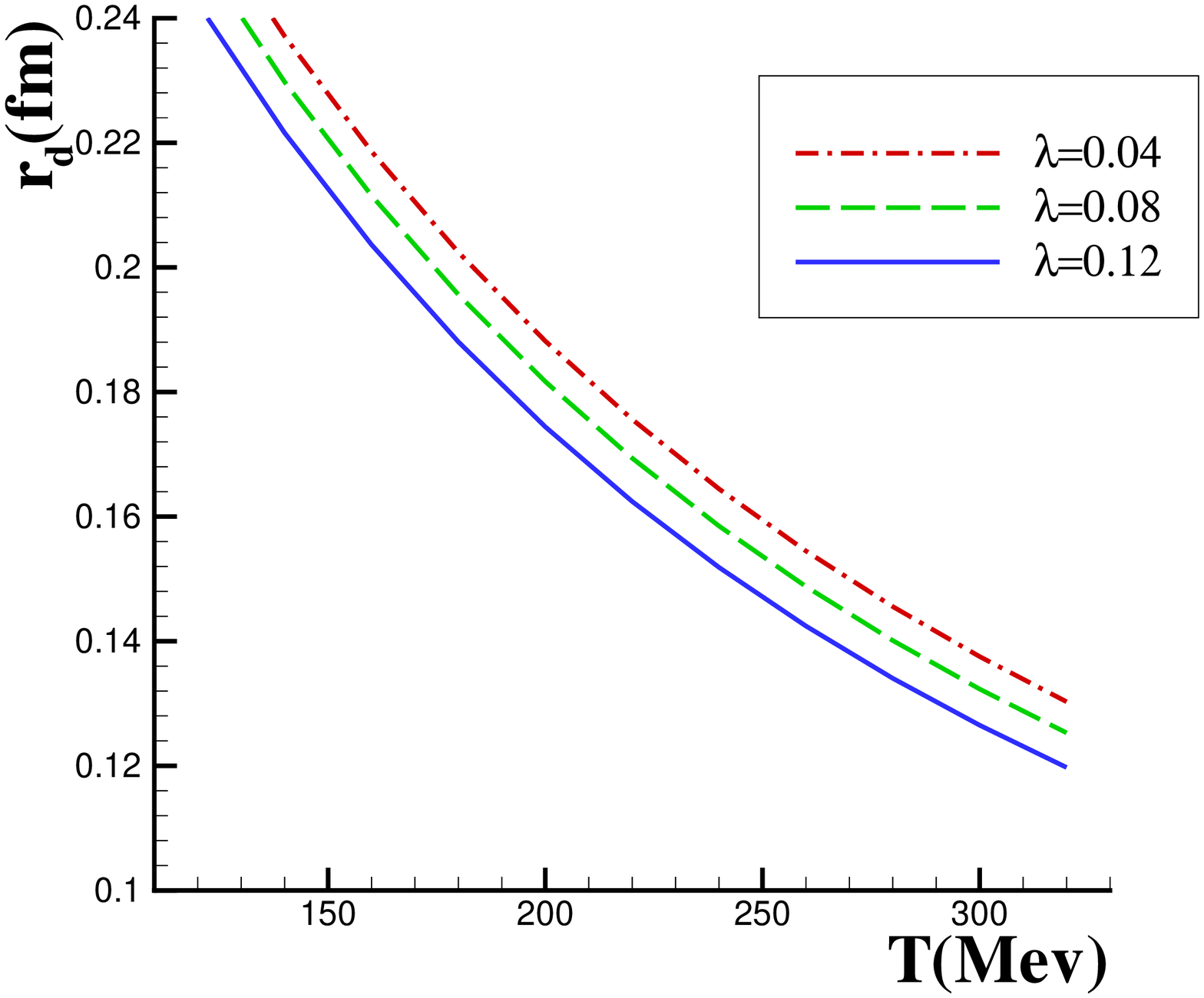}}%
\caption{Left: The inter-quark distance versus the chemical
potential in the quark-gluon plasma at fixed temperature $T=100MeV$
and $n=1$. Right: The inter-quark distance versus the temperature in
the quark-gluon plasma at fixed $\mu=30 MeV$ and $n=1$. }
\end{figure}%

We intend to study the effect of the higher derivative corrections
to the heavy quark potential \eqref{V} and the inter-quark distance
\eqref{r}. We cannot solve \eqref{V} and \eqref{r} analytically and
we have to resort to numerical methods. For different values of
$\lambda= 0.04, 0.08$ and $0.12$, the binding energy of the heavy
meson versus the inter-quark distance is plotted in the Fig. 4. It
is clearly seen that at a given inter-quark distance by increasing
the $\lambda_{}$ the binding energy also increases. This is an
interesting phenomena, because it implies that as the coupling
constant increases the U-shaped of the connected string may be
discount. We leave physics of this phenomena for further
investigation.

At a given energy, the increase in the $\lambda_{}$ leads to a
decrease in the $r_d$. This confirms that considering the higher
derivative corrections in the gravity background leads to a decrease
in the dissociation length of heavy meson in the boundary gauge
theory \cite{Fadafan:2011gm}. However in our study, the boundary
gauge theory consists of the light and heavy quarks. Then one can
extend the results of \cite{Fadafan:2011gm} to the medium quark,
too. This result has been observed for the first time in this study.
To more clarify this observation, we plot the dissociation length
$r_d$ versus the chemical potential and the temperature of the hot
plasma in Fig. 5. In this figure, we have fixed the coupling
constant as $\lambda_{}=0.04,.08$ and $0.12$. One finds from this
figure that at a given temperature or chemical potential, by
increasing the $\lambda_{}$, $r_d$ becomes shorter. The current
lattice results show the surprising resistance to the melting of the
1S states such as $J/\psi$, $\eta_c$ at least up to $T > 1.5 T_c$.
The $\psi'(2S)$ and $\chi_c(1P)$, however, do melt right above $T_c$
even in the lattice determination \cite{Satz:2009hr}. One finds that
by considering higher derivative corrections the heavy meson in the
medium composed of light quarks and gluons will melt at higher
temperature.

\section{The rotating heavy meson in the quark medium}
It is desirable to take into account the nonzero angular momentum of
the quark antiquark pair in our study. Excited $q-\bar{q}$ pair
could have the non-zero angular momentum and spin.  A holographic
picture of the melting process of low spin mesons in the quark-gluon
plasma was discussed in \cite{Hoyos:2006gb}. The large spin case was
also studied in \cite{Peeters}.

In this section, we are going to obtain some information for an
exited heavy $q-\bar{q}$ pair in plasma with a non-zero chemical
potential from the gravity dual theory. We consider the rotating
heavy meson in the quark gluon plasma phase of QCD and investigate
melting of it. We expect that excited heavy mesons like $\psi'(2S)$
and $\chi_c(1P)$ have a larger dissociation length than the ground
state mesons like $J/\psi$. It is shown that by increasing the spin
of heavy meson, the dissociation length also increases which
confirms our expectation. This subject was also studied in
\cite{Peeters,Antipin,AliAkbari:2009pf}. Here, we discuss the
inter-quark distance in the presence of the chemical potential.

We consider a rotating meson on the $\rho,\theta$ plane with $x_3$
the direction perpendicular to the plane of rotation. Then the space
time metric in \eqref{ds2}, will be as
\begin{eqnarray}\label{adsschmetric}
 ds^{2}=\frac{R^2}{z^2}\big(f(z)dt^{2}+ d\rho^{2}+\rho^{2}d
 \theta^{2}+dx_{3}^{2}+\frac{dz^{2}}{f(z)}\big),
\end{eqnarray}
where our four-dimensional space is along $t,\rho,\theta$ and $x_3$.
In order to describe the rotation of a quark-antiquark pair the
end-points of the string on the probe brane must satisfy the Neumann
boundary conditions.
To study a rotating string, we make use of the Nambu-Goto action in
the above background given by
\begin{eqnarray}
S=-\frac{1}{2\pi\alpha'}\int d\tau d\sigma\sqrt{-{\rm{det}}g_{ab }
}.
\end{eqnarray}
The coordinates $(\sigma, \tau)$ parameterize the induced metric
$g_{ab}$ on the string world-sheet. Indices $a,b$ run over the two
dimensions of the world-sheet. Let $X^\mu(\sigma, \tau)$ be a map
from the string world-sheet into spacetime and let us define $\dot X
=\partial_\tau X$, $X' =
\partial_\sigma X$, and $V \cdot W = V^\mu W^\nu G_{\mu\nu}$ where
$G_{\mu\nu}$ is the AdS black hole metric. Indices $\mu, \nu$ run
over the five dimensions of spacetime. Then
\begin{eqnarray}
 -g=-{\rm{det}}g_{ab }=(\dot X \cdot X')^2 - (X')^2(\dot X)^2.
\end{eqnarray}%
We choose to parameterize the two-dimensional world-sheet of the
rotating string
$X^\mu(\sigma, \tau)$ according to %
\be\label{ansatz}
 X^\mu(\sigma, \tau)=\left(t=\tau,\,\,\,\rho=\sigma,\,\,\, z=z(\rho),\,\,\,\theta=\omega t\right).
\ee %

Simply, what this means is that the radius of the rotating
quark-antiquark on the probe brane changes with the fifth direction
of the bulk space as we move further into the bulk. In arriving at
the parametrization (\ref{ansatz}), we made use of the fact that the
quark-antiquark pair is in circular motion with radius $d$ at a
constant angular velocity $\omega$. Also, we assumed that the system
retains its constant circular motion at all times. Furthermore, the
ansatz (\ref{ansatz}) does not show any dragging effects which frees
us from applying a force to maintain the rigid rotation
\cite{Peeters}.

According to our ansatz (\ref{ansatz}), the Nambu-Goto action becomes%
\begin{eqnarray}
 S=-\frac{R^2}{2\pi\alpha'}\int dt\,d\rho\frac{1}{z^2}\sqrt{\bigg(1-mz^4+q^2z^6-\rho^2\omega^2 \bigg)
 \left(\frac{z'^2}{1-mz^4+q^2z^6}+1\right)  },\label{action}
\end{eqnarray}
where prime is the derivative with respect to $\rho$. It is evident
that positivity of the square root in \eqref{action} requires that
$1-mz^4+qz^6-\rho^2 \omega^2 \geq 0$. In the case of zero finite
density, it can be verified that all rotating strings with different
angular velocity do satisfy this condition \cite{AliAkbari:2009pf}.
As discussed by \cite{CasalderreySolana:2007qw} at this radial
coordinate a horizon develops on the world-volume. The stochastic
trailing string and Langevin dynamics was studied in
\cite{Giecold:2009cg}. It was shown that the stochasticity arises at
the string world-sheet horizon, and thus is causally disconnected
from the black hole horizon in the space-time metric. They conclude
that this hints at the non-thermal nature of the fluctuations. It
would be interesting to extend the results to the case of finite
density, {\em i.e.} $RNAdS$ black hole. We leave this problem for
further work.

We define lagrangian density as%
\begin{eqnarray}
\mathcal{L}=\frac{1}{z^2}\sqrt{\bigg(1-mz^4+q^2z^6-\rho^2\omega^2
\bigg)
 \left(\frac{z'^2}{1-mz^4+q^2z^6}+1\right)  },\label{lag}
\end{eqnarray}
and the equation of motion for $z$ is given by
\begin{eqnarray}\label{max4}
 \partial_\rho \left(  \frac{z'\sqrt{\bigg(1-mz^4+q^2z^6-\rho^2\omega^2
\bigg)
 \left(\frac{z'^2}{1-mz^4+q^2z^6}+1\right)  }}{z^2\,(z'^2+1-mz^4+q^2z^6)} \right) -
 \frac{\partial {\cal{L}}}{\partial z}=0,\label{EOM}
\end{eqnarray}

where $\frac{\partial {\cal{L}}}{\partial z}$ is easily found from \eqref{lag}. %
The equation of motion for $z(\rho)$ in (\ref{EOM}) is nonlinear and
coupled. However, for generic values of $\omega$ we cannot solve
this equation analytically and we have to resort to numerical
methods. The boundary conditions which solve \eqref{max4} physically
means that string terminates orthogonally on the brane in the
boundary which in turn implies Neumann boundary conditions. Also
$z'(0)=0$ has been considered at the tip of the string where
$\rho=0$. To check the validity of our solutions, we choose $\rho=d$
so that at $\rho=0$ we keep the condition $z'(0)=0$. This system was
studied in details in  \cite{AliAkbari:2009pf}. It was argued that
as $\omega$ decreases the string endpoints become more and more
separated, {\em i.e.} the radius of the open string at the boundary
increases and it penetrates deeper into the horizon. We also find
the same physics in our case.
\begin{figure}[ht]
\centerline{\includegraphics[width=3in]{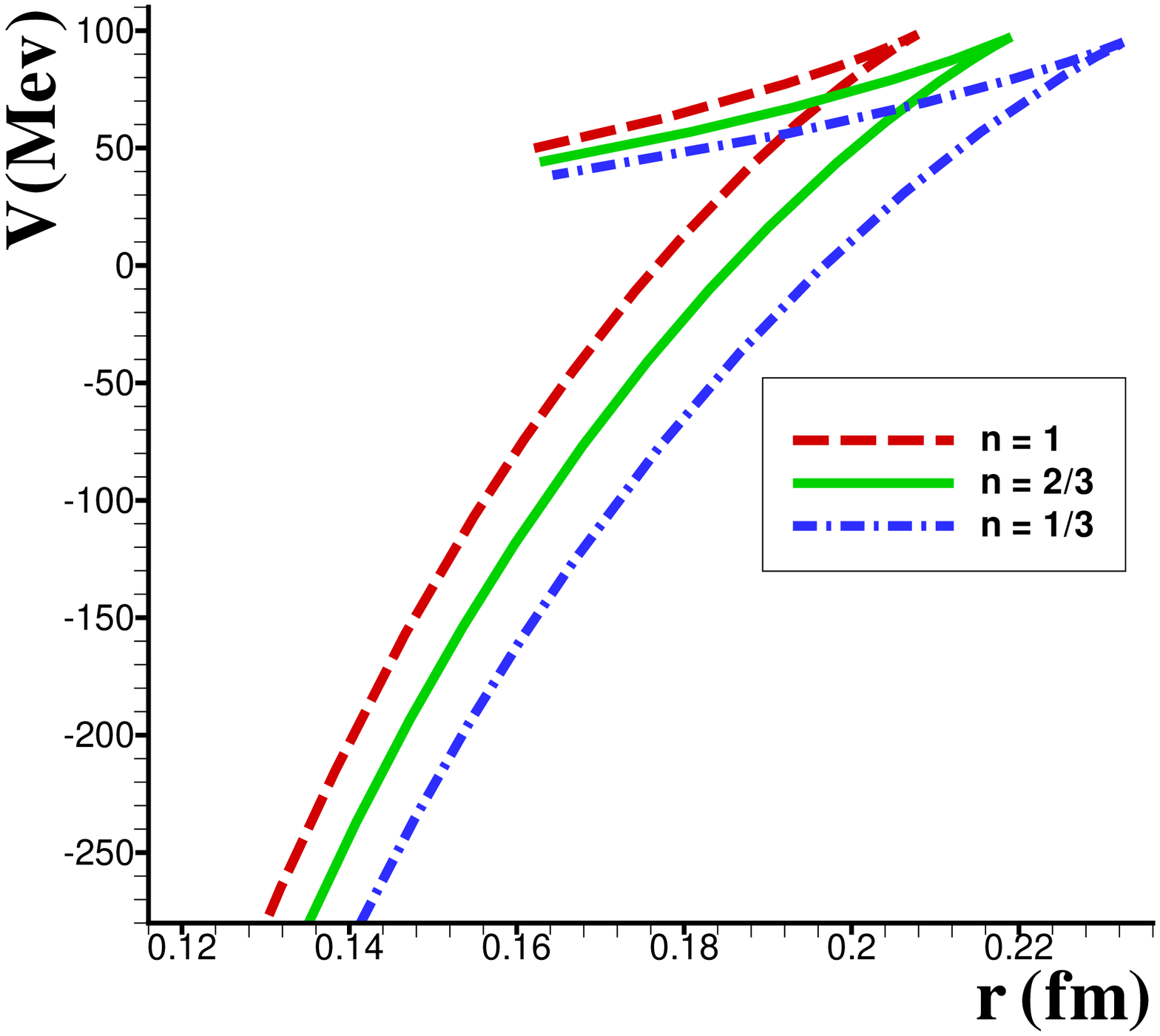}\includegraphics[width=3in]{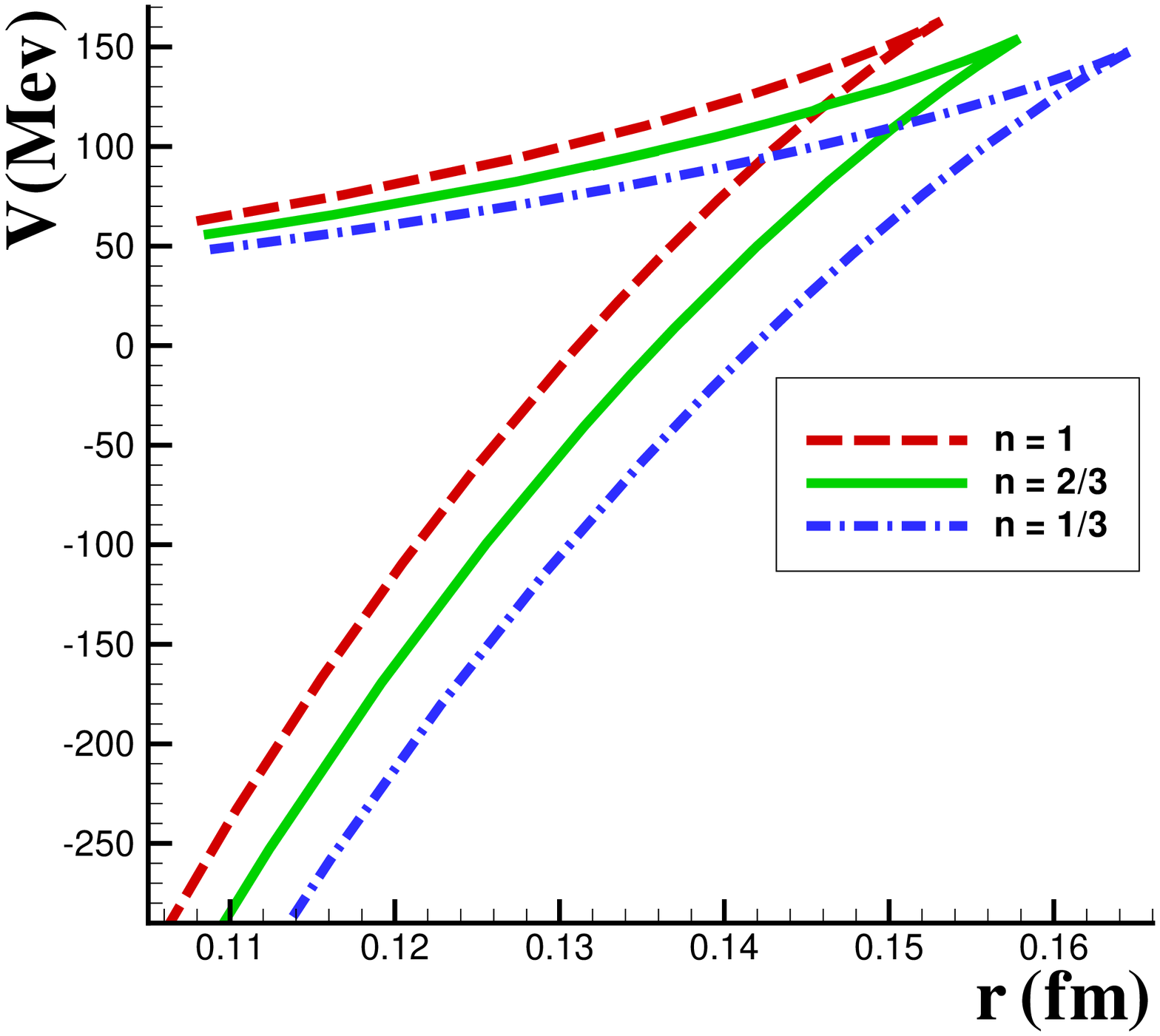}}%
\caption{Left: The binding energy of a rotating heavy meson
 versus the inter-quark distance in the
quark-gluon plasma at fixed $\mu=30MeV$ and $T=200MeV$.
Left:$\omega=500$. Right: $\omega=1500$. }
\end{figure}%
The constant of the motion is angular momentum
$J$ which can be found as\\%
\be J=\frac{\rho^2 \omega}{z^2} \sqrt{\frac{1+z'^2/f(z)}{f(z)-\rho^2\omega^2}}\ee%
\\%
At the tip of the string, we have%
\be z=z_0,\,\,\,\,\,\rho=0.\ee%
It is clear that at the tip of the string $J=0$. This is reasonable,
because this is the special point on the string which is static all
the times.

\begin{figure}[ht]
\centerline{\includegraphics[width=3in]{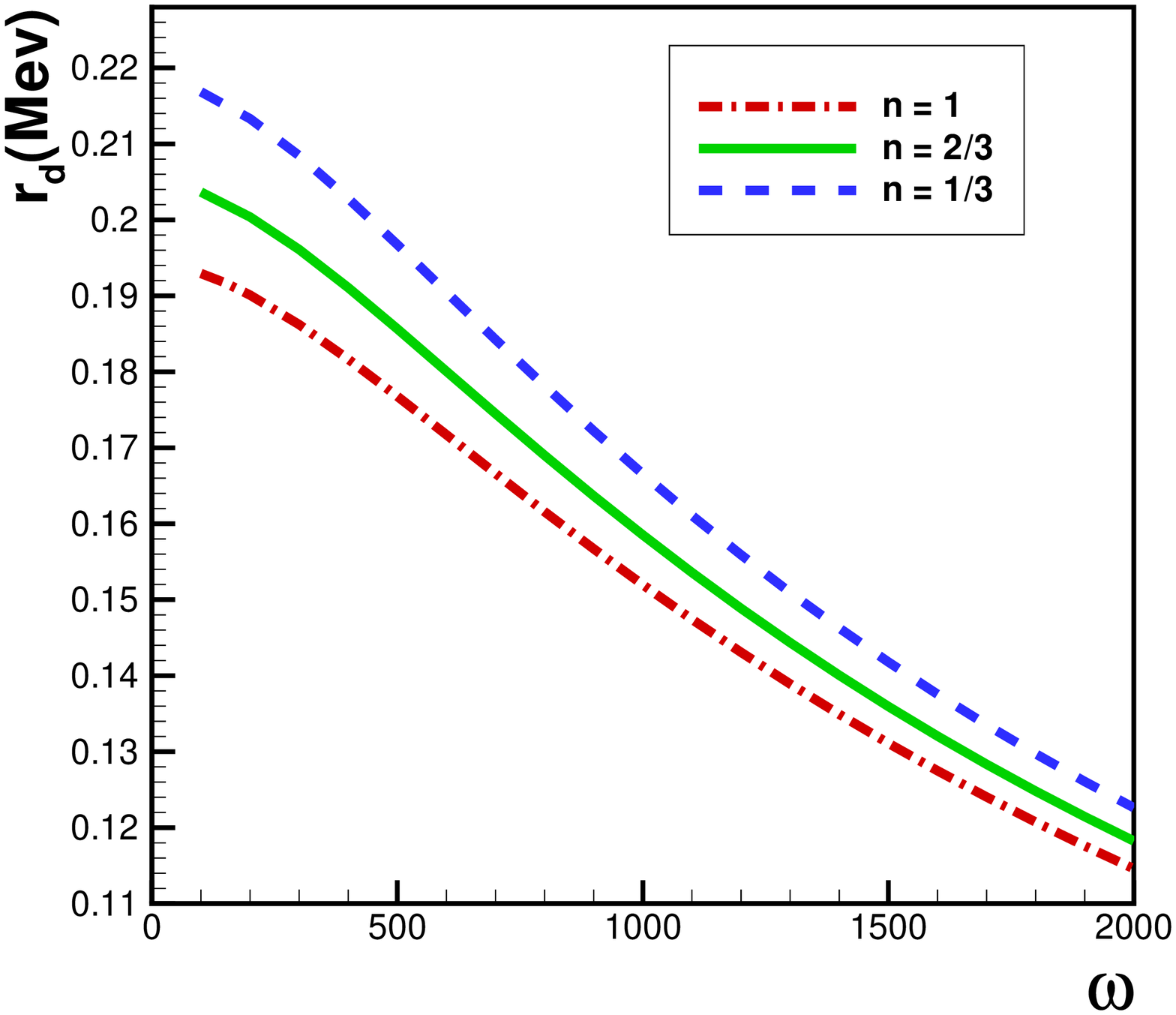}\includegraphics[width=3in]{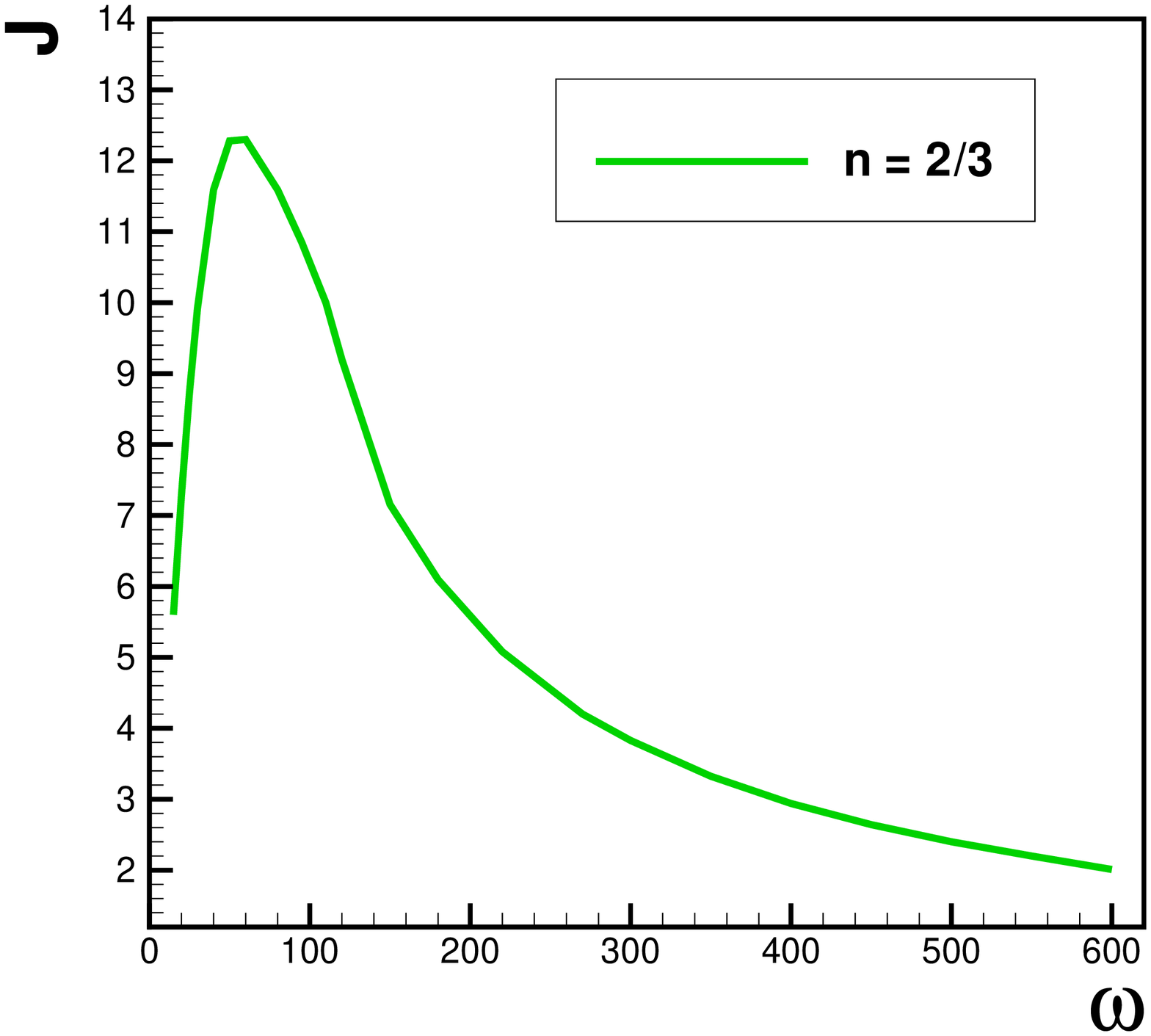}}%
\caption{Left: The inter-quark distance versus the angular-momentum
in the quark-gluon plasma phase at fixed $\mu=30MeV$ and $T=200MeV$.
Right: The spin of heavy meson versus the angular-momentum. }
\end{figure}%

It would be interesting to investigate melting of a rotating heavy
meson in the quark medium.\footnote{We restrict our attention to the
case of quark-gluon phase.} Then one should study the binding energy
of the rotating meson. One finds the standard calculations in
\cite{Antipin,AliAkbari:2009pf}. We present the numerical results in
the Fig. 6. We have fixed the chemical potential and the temperature
as $\mu=30MeV$ and $T=200MeV$, respectively. One finds that as the
angular-momentum  $\omega$ increases, the binding energy also
increases. Also as the static case, by increasing the flavor number
$N_f=nN_c$ the inter-quark distance becomes shorter. To more clarify
the effect of the $\omega$ on the inter-quark distance, we plot
$r_d$ versus the $\omega$ in the left plot of Fig. 7. One concludes
that the maximum height of the radius of the quark-antiquark system
depends on angular velocity.

One finds from this figure that by decreasing $\omega$, the
dissociation length becomes larger. What is the relation between
angular-momentum and spin of particle? The right plot of Fig. 7
explores this question. The physical part is when by decreasing the
angular momentum, the spin of particle increases \cite{Peeters}. The
other part is related to the long strings which are not stable. As a
result we find that by increasing the spin of heavy meson, the
dissociation length also increases which confirms our expectation
from lattice. The same physics was discussed in the case of rotating
baryons in \cite{baryon}, too.

\section{Conclusion}
The melting of quarkonium states is of high importance in the
physics of QGP at RHIC and LHC. It has long been regarded to be one
of the cleanest signatures of plasma formation. In particular,
quarkonium states such as the $J/\psi$ meson are expected to melt in
the QGP at higher temperatures than the excited states such as
$\psi'(2S)$ and $\chi_c(1P)$.

In this paper we have studied the dissociation length of a heavy
meson in the presence of chemical potential. In particular, the
inter-quark distance in the static and rotating case was
investigated. It was shown that in both phase of QCD, i.e
quark-gluon plasma and hadronic phase the dissociation length of
quark-antiquark pair becomes shorter as the temperature or the
chemical potential increases. We conclude that the melting mechanism
in the quark-gluon plasma and in the hadronic phase are the same,
{\em{i.e}}. the interaction between heavy quarks is screened by the
light quarks.

We have considered the static case and extended the results to
higher derivative corrections, {\em{i.e}}. ${\cal{R}}^2$ which
correspond to finite coupling corrections in the hot plasma. As one
expects from the results of \cite{Fadafan:2011gm}, the dissociation
length becomes shorter as the coupling constant increases. This
confirms the claim of \cite{Fadafan:2011gm} and extends it to the
medium composed of light quarks and gluons.

To study melting of excited quarkonium states, the dissociation
length of a rotating heavy meson in the quark-gluon plasma phase of
QCD was also studied. It was shown that by increasing the spin of
heavy meson, the dissociation length also increases which confirms
our expectation from Lattice. Also the effect of increasing flavor
quarks on the dissociation length was investigated and it was shown
that as the flavor quarks increase, the dissociation length
decreases.

\section*{Acknowledgment}
We would like to thank H. Niazy, C. Park and S. Sheikh-Jabbari for
useful
discussions and M. Sohani for reading the manuscript, carefully.\\ \\%
\textbf{Appendix A:}\\%
In this appendix, we calculate the integral in \eqref{rz00}. We
consider the  general formula as%
\be \int_y^\infty \frac{dx}{\sqrt{(x-a)(x-b)(x-c)}}= g
F(\phi,k),\label{Integral}
\ee%
where%
\be
g=\frac{2}{\sqrt{a-c}}\,,\,\,\,\,\,\,\,\,k^2=\frac{b-c}{a-c}\,,\,\,\,\,\,\,\,
\phi=Sin^{-1}\sqrt{\frac{a-c}{y-c}.}\ee%
To calculate \eqref{rz00}, we convert it to the form of
\eqref{Integral}. We change variable $z$ as%
\be z=\frac{1}{\sqrt{x}}.\ee%
Then one should solve this integral%
\be \int_{\frac{1}{z_0^2}}^{\infty} \frac{dx}{\sqrt{-q^2+m x-x^3}},\ee%
The parameters of $a,b $ and $c$ can be found by solving these
equations:%
\be a+b+c=0,\,\,\,\,\,\,\\
abc=-q^2,\,\,\,\,\,\, (ab+ac+bc)=-m.
 \ee%


\end{document}